\documentclass{article}
\usepackage{graphicx}
\usepackage{bm}
\usepackage{amsmath}
\usepackage{amsfonts}
\usepackage{amssymb}
\usepackage[utf8]{inputenc}
\usepackage[american]{babel}

\newcommand{\ket}[1]{|#1\rangle}
\newcommand{\bra}[1]{\langle#1|}
\newcommand{\braket}[3]{\langle#1|#2|#3\rangle}
\newcommand{\scalar}[2]{\langle#1|#2\rangle}
\newcommand{\op}[1]{|#1\rangle\langle#1|}

\begin{document}

\title{Spatial quantum search in a triangular network}

\author{G. Abal$^1$\thanks{corresponding author: abal@fing.edu.uy}, R. Donangelo$^1$, M. Forets$^1$ and R.~Portugal$^2$\\
{\small $^1$Instituto de F\'{\i}sica, Facultad de Ingenier\'{\i}a, UdelaR,} \\ 
{\small C.C. 30, C.P. 11300, Montevideo, Uruguay} \\
{\small $^2$Laborat\'{o}rio Nacional de Computa\c{c}\~{a}o Cient\'{\i}fica - LNCC,} \\ 
{\small Av.  Get\'{u}lio Vargas 333, Petr\'{o}polis, RJ, 25651-075, Brazil}}




\maketitle

\begin{abstract}
The spatial search problem consists in minimizing the number of steps required to find a given site in a
network, under the restriction that only oracle queries or translations to neighboring sites are allowed. We propose a quantum algorithm for the spatial search problem on a triangular lattice with $N$ sites and
torus-like boundary conditions. The proposed algorithm is a special case of the general framework for abstract search proposed by Ambainis, Kempe and Rivosh~\cite{AKR05} (AKR) and Tulsi~\cite{Tulsi08}, applied to a triangular network. The AKR-Tulsi formalism was employed to show that the time complexity of the quantum search on the triangular lattice is $O(\sqrt{N \log N})$.
\end{abstract}

\section{Introduction}

The spatial quantum search problem consists of using local unitary operations to search for one (or more) nodes within a set of $N$ spatially arranged sites with an implicit notion of distance between them. The search nodes are identified by the non-zero values of a binary function (the oracle), as usual. The spatial search problem~\cite{AA05} incorporates the restriction that, in one step, one can either query the oracle at the current site or advance to a neighboring site. It has been pointed out by Benioff~\cite{Benioff} that in a two-dimensional network under this restriction, Grover's search~\cite{Grover},~\cite{Grover2} provides no advantage in terms of running time over a classical search due to the intrinsic non-locality of Grover's symmetrization. Ambainis et al. proposed a generalized formalism for quantum walk (QW) based spatial search algorithms and worked out the specific case of a two-dimensional cartesian network, obtaining a $O(\sqrt{N}\log N)$ algorithm~\cite{AKR05}, to which we shall refer to as AKR. Tulsi has proposed an improvement to AKR which requires an ancilla qubit and leads to an $O(\sqrt{N\log N})$ algorithm in two dimensions~\cite{Tulsi08}. However, it is not known whether the optimal solution is $O(\sqrt{N})$ for the two-dimensional spatial search problem. In contrast, it is known that the optimal solution is achieved in higher dimensions, such as 3D-grid~\cite{AKR05} and the SKW algorithm~\cite{SKW03}, which searches an item within $N=2^n$ sites arranged in an $n$--dimensional $(n>2)$ hypercube.

There are three ways to cover the plane with regular polygons: squares, hexagons and triangles. The resulting regular networks differ in their degree $d$ (number of connections per node) which is $4, 3$ and $6$ respectively. As mentioned before, for the rectangular grid $(d=4)$ the AKR search algorithm finds a marked vertex in time $O(\sqrt{N}\log N)$.  A QW-based spatial search algorithm of time complexity $O(\sqrt{N}\log N)$  has recently been implemented for a two-dimensional hexagonal network $(d=3)$~\cite{Hexagons}. Both algorithms can be improved to $O(\sqrt{N\log N})$ with Tulsi's modification. These results suggest that the degree or connectivity of a regular network does not affect the performance of a QW-based search algorithm.

In this work, a new search algorithm for the case of a triangular network is proposed and analyzed. The proposed algorithm is a special case of the general framework for abstract search~\cite{AKR05, Tulsi08}, so we employ the AKR-Tulsi formalism to show that the time complexity of the quantum search on the triangular lattice is $O(\sqrt{N \log N})$. This provides further evidence that the degree of the underlying network does not affect the performance of the quantum algorithm.

The paper is organized as follows. In Section \ref{sec:QW3} we discuss the implementation of a QW on the triangular network. In Section \ref{sec:complexity} we analyze the time complexity of this search algorithm. In Section \ref{sec:simulation} we perform a numerical analysis of the algorithm. Finally in Section \ref{sec:final} we present our conclusions.

\section{QW on a triangular Network}
\label{sec:QW3}

\begin{figure}
  \centering
  \includegraphics[scale=0.4]{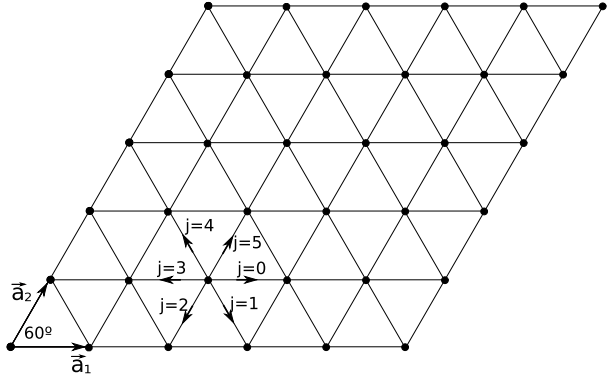}
  \caption{\footnotesize A $\sqrt{N}\times \sqrt{N}$ triangular network (here $N=36$). The $N$ sites form a Bravais lattice and the six directions of motion are labelled by an integer $j\in [0,5]$.}
  \label{fig:lattice}
\end{figure}

Let us consider $N$ sites arranged in a triangular network covering a two-dimensional region, as shown in Figure \ref{fig:lattice}. The network is $\sqrt{N}\times \sqrt{N}$ and periodic boundary conditions are assumed. A site on the lattice is located by two integers $(n_1,n_2)$ according to
\begin{equation}
 \mathbf{r}=n_1 \mathbf{a}_1 + n_2 \mathbf{a}_2,
\end{equation}
where $\mathbf{a}_1$ and $\mathbf{a}_2$ are unit vectors forming a $60^o$ angle, as indicated in Figure~\ref{fig:lattice}. These integers are such that $n_i\in [0,\sqrt{N}-1]$ for $i=1,2$ and thus each of them takes $\sqrt{N}$ different values.

These sites define an orthonormal set of quantum state vectors, $\{\ket{n_1,n_2}\}$, which span an $N$--dimensional Hilbert space ${\cal H}_P$. At a given site, there are six possible directions of motion which we label with an integer $j\in [0,5]$, as indicated in Figure~\ref{fig:lattice}. The orthonormal states $\{\ket{j}\}$ span a six-dimensional Hilbert space, ${\cal H}_C$, which we shall refer to as the ``coin" subspace. The Hilbert space for this problem, $\cal H={\cal H}_C\otimes {\cal H_P}$, is $6N$--dimensional. A generic state vector is expressed as
\begin{equation}
 \ket{\Psi}=\sum_{j=0}^{5} \sum_{n_1, n_2=0}^{\sqrt{N}-1} a_{j,\hat{n}}\, \ket{j,\hat{n}}, \label{eq:state_vector}
\end{equation}
where the $a_{j,\hat{n}}$ are complex amplitudes which satisfy the normalization constraint and we have introduced the shorthand notation $\hat n\equiv (n_1,n_2)$.

The standard QW on this network is implemented with a unitary evolution operator of the form
\begin{equation}\label{eq:U}
U=S\cdot(C\otimes I_P)
\end{equation}
where $S$ is a shift operator in $\cal H$ (to be specified below), $I_P$ is the identity  operation in ${\cal H}_P$ and $C$ is a unitary coin operation in ${\cal H}_C$. The useful coin operation for spatial search problems~\cite{AKR05,SKW03} is Grover's coin, whose matrix elements for a $d$--dimensional space are $G^{(d)}_{ij}=\frac{2}{d}-\delta_{ij}$. For the particular case $d=6$, it is given by
\begin{equation}
G^{(6)}=   \frac{1}{3} \left( \begin{array}{cccccc}
      -2 & 1 & 1 & 1 & 1 & 1 \\
1 & -2 & 1 & 1 & 1 & 1 \\
1& 1 & -2 & 1 & 1 & 1 \\
1 & 1 & 1 & -2 & 1 & 1 \\
1 & 1 & 1 & 1 & -2 & 1 \\
1 & 1 & 1 & 1 & 1 & -2 \\
\end{array}
    \right).
\end{equation}
Thus we use $C=G^{(6)}$. The shift operator implements single-step displacements acting on the kets $\ket{j;n_1,n_2}$ in the form
\begin{eqnarray}
 S\ket{0; n_1, n_2}&=&\ket{3; n_1+1,n_2}, \nonumber\\
  S\ket{1; n_1, n_2}&=&\ket{4; n_1+1,n_2-1}, \nonumber\\
 S\ket{2; n_1, n_2}&=&\ket{5; n_1,n_2-1}, \nonumber\\
 S\ket{3; n_1, n_2}&=&\ket{0; n_1-1,n_2}, \label{eq:Sr}\\
 S\ket{4; n_1, n_2}&=&\ket{1; n_1-1,n_2+1}, \nonumber\\
 S\ket{5; n_1, n_2}&=&\ket{2; n_1,n_2+1}.\nonumber
\end{eqnarray}
Note that $S$ inverts the coin state. This invertion is crucial for the efficiency of the search algorithm described in the next section. Finally, the dynamics of the QW is obtained by applying $U$ repeatedly $\ket{\Psi(m)}=U^m\ket{\Psi(0)}$ for some integer $m$.

The standard QW is best analyzed in the Fourier-transformed space. Let us consider the reciprocal lattice vectors $\{ \mathbf{g}_1, \mathbf{g}_2\}$, which satisfy the usual requirements from condensed matter physics~\cite{Kittel}
\begin{equation}
 \mathbf{g}_i \cdot \mathbf{a}_j = \frac{2\pi}{\sqrt{N}} \delta_{ij}.
\end{equation}
A site in this \textit{reciprocal lattice} is located by
\begin{equation}
 \mathbf{k}=k_1  \mathbf{g}_1 + k_2 \mathbf{g}_2,
\end{equation}
with $k_1, k_2$ integers in $[0,\sqrt{N}-1]$. Let us use the notation $\hat{k}=(k_1,k_2)$ and write a generic state vector in the Fourier representation as
\begin{equation}
 \ket{\Psi}=\sum_{j,\hat{k}} f_{j,\hat{k}} \ket{j,\hat{k}}.
\end{equation}
The kets $\ket{j,\hat{k}}$ and $\ket{j,\hat{n}}$ are related by the discrete Fourier transform
\begin{eqnarray}
\ket{\hat{k}}&=&\frac{1}{\sqrt{N}} \sum_{\hat{n}} \textrm{e}^{-i\mathbf{k}\cdot\mathbf{r}} \ket{\hat{n}}\label{eq:hat_k}\\
\ket{\hat{n}}&=& \frac{1}{\sqrt{N}} \sum_{\hat{k}} \textrm{e}^{i\mathbf{k}\cdot\mathbf{r}} \ket{\hat{k}}
\end{eqnarray}
and one can check that $\scalar{\hat{k}}{\hat{n}}=\omega^{\hat n\cdot \hat k}/\sqrt{N}$ where $\omega\equiv \textrm{e}^{2\pi i/\sqrt{N}}$ and $\hat n\cdot \hat k = n_1k_1 + n_2k_2$. The action of $S$ on the kets $\ket{j,\hat{k}}$ of the Fourier representation can be obtained from Eqs.~(\ref{eq:Sr}) as
\begin{equation}
\begin{array}{rclcrcl}
 S\ket{0,\hat{k}}&=&w^{k_1}\ket{3,\hat{k}},&\quad& S\ket{1,\hat{k}}&=&w^{k_1-k_2}\ket{4,\hat{k}},\\
 S\ket{2,\hat{k}}&=&w^{-k_2}\ket{5,\hat{k}}, &\quad& S\ket{3,\hat{k}}&=&w^{-k_1}\ket{0,\hat{k}},\\
 S\ket{4,\hat{k}}&=&w^{-k_1+k_2}\ket{1,\hat{k}},&\quad&  S\ket{5,\hat{k}}&=&w^{k_2}\ket{2,\hat{k}}.\label{eq:Sk}
 \end{array}
\end{equation}
Thus, in the $\hat k$--representation $S$ acts diagonally, i.e. $S=\sum_{\hat k} S_k\op{\hat k}$, where $\protect{S_k=\braket{\hat k}{S}{\hat k}}$ is the reduction of $S$ to ${\cal H}_C$. Therefore, the evolution operator, Eq.~(\ref{eq:U}), is also diagonal in the Fourier representation and can be expressed as $U=\sum_{\hat k} U_{\hat k}\op{\hat k}$, where $U_{\hat k}=\braket{k}{U}{k}$ acts in ${\cal H}_C$. The matrix elements of the reduced operator can be calculated from Eq.~(\ref{eq:Sk}), with the result
\begin{equation}
U_{\hat{k}}=   \frac{1}{3} \left( \begin{array}{cccccc}
      w^{-k_1} & w^{-k_1} & w^{-k_1} & -2w^{-k_1} & w^{-k_1} & w^{-k_1} \\
      w^{-k_1+k_2} & w^{-k_1+k_2} & w^{-k_1+k_2} & w^{-k_1+k_2} & -2w^{-k_1+k_2} & w^{-k_1+k_2} \\
      w^{k_2} & w^{k_2} & w^{k_2} & w^{k_2} & w^{k_2} & -2w^{k_2} \\
      -2w^{k_1} & w^{k_1} &  w^{k_1} & w^{k_1} & w^{k_1} & w^{k_1} \\
      w^{k_1-k_2} & -2 w^{k_1-k_2} & w^{k_1-k_2} &  w^{k_1-k_2} & w^{k_1-k_2} & w^{k_1-k_2} \\
      w^{-k_2} & w^{-k_2} & -2 w^{-k_2} & w^{-k_2} & w^{-k_2} &  w^{-k_2} \\
\end{array}
    \right).
\end{equation}

The characteristic polynomial factors as
\begin{equation}\label{eq:Pchar}
P(\lambda)=(\lambda-1)^2(\lambda+1)^2(\lambda^2-2\cos(\theta_k)\lambda+1),
\end{equation}
where $\theta_k$ is defined by
\begin{equation}\label{cos_theta}
 \cos(\theta_k) \equiv \frac{1}{3} \left( \cos(\tilde{k_1}) + \cos(\tilde{k_2}) + \cos(\tilde{k_1}-\tilde{k_2}) \right),
\end{equation}
and $\tilde{k}_i \equiv \frac{2\pi k_i}{\sqrt{N}}$ for $i=1,2$. The eigenvalues are $\pm 1$, each with multiplicity 2, and $\textrm{e}^{\pm i\theta_k}$. Let us denote by $\ket{\nu_{\pm 1}}$ and $\ket{\nu_{\pm k}}$ the normalized eigenvectors of $U_{\hat{k}}$ associated with the eigenvalues $\pm 1$ and $\textrm{e}^{\pm i\theta_k}$, respectively. Then  $\ket{\nu_{\pm 1},\hat k}$ and $\ket{\nu_{\pm k},\hat k}$ are eigenvectors of $U$ associated with the same eigenvalues, where $\ket{\hat k}$ is defined in Eq.~(\ref{eq:hat_k}).

\section{Time complexity of the search algorithm}
\label{sec:complexity}

We shall use Tulsi's version~\cite{Tulsi08} of the framework of the \textit{abstract search algorithm}~\cite{AKR05} to analyze the time complexity of a search algorithm on the triangular network. We assume that there is a single marked vertex $\ket{\hat t}$, which we want to find. The search algorithm uses a conditional coin operation, which acts as $-I_C$ on the searched site $\ket{\hat t}$ and as $G^{(6)}$ otherwise. Thus, the modified evolution operator is $U'=S\cdot C'$, where $C'$ acts in $\cal H$ as just described, i.e.
\begin{equation}\label{coin_operator}
    C' =  -I_C\otimes \op{\hat t} +\sum_{\hat n\neq \hat t} G^{(6)}\otimes\op{\hat n}.
\end{equation}
AKR~\cite{AKR05} have shown that the evolution of the modified quantum walk $U'$ may be analyzed using the eigenspectrum of the standard QW evolution operator $U$. This fact actually reduces the analysis of the search algorithm to a tractable eigenproblem for the unitary operator $U$.

\begin{figure}[h]
  \centering
  \setlength{\unitlength}{0.65pt}
  \begin{picture}(160,160)(80,0)
    \put(43,25){\makebox(0,0)[r]{$\vert{u_{P}}\rangle$}}
    \put(250,125){\line(1,0){20}}
    \put(80,110){\framebox(30,30){$X_{\delta}$}}
    \put(140,110){\framebox(30,30){$X_{\delta}^{\dagger}$}}
    \put(220,110){\framebox(30,30){$-Z$}}
    \put(218,25){\line(1,0){52}}
    \put(218,75){\line(1,0){52}}
    \put(102,15){\framebox(46,70){\small{$R_{\hat t}$}}}
    \put(148,25){\line(1,0){24}} \put(148,75){\line(1,0){24}}
    \put(195,125){\circle*{8}} \put(195,121){\line(0,-1){36}}
    \put(172,15){\framebox(46,70){$U$}}
    \put(43,125){\makebox(0,0)[r]{$\vert{1}\rangle$}}
    \put(43,75){\makebox(0,0)[r]{$\vert{u_{C}}\rangle$}}
    \put(125,125){\circle*{8}}
    \put(170,125){\line(1,0){50}}
    \put(110,125){\line(1,0){30}}
    \put(55,125){\line(1,0){25}}
    \put(125,121){\line(0,-1){36}}
    \put(55,25){\line(1,0){47}}
    \put(55,75){\line(1,0){47}}
  \end{picture}
  \caption{Tulsi's circuit diagram for the one-step evolution operator of the quantum walk search algorithm.}\label{fig:circ}
\end{figure}
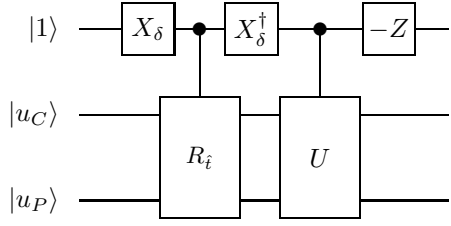

The evolution operator $U'$ can be written in another form, useful for Tulsi's modified algorithm. This modification requires an extra register (an ancilla qubit) used as a control for the operators $R_{\hat t}$ and $U$. Using Eq.~(\ref{coin_operator}), one can show that $U'=U\cdot R_{\hat t}$~, where $R_{\hat t}=I_{6N}-2\ket{u_C,\hat t}\bra{u_C,\hat t}$, $U$ is given by Eq.~(\ref{eq:U}) and $\ket{u_C}$ is the uniform superposition of the computational basis of the coin space. The operators acting on the ancilla register are described in Figure~\ref{fig:circ}, where $-Z$ is the negative of Pauli's $Z$ operator
and
\begin{equation}
  X_{\delta}= \left( \begin{array}{cc} \cos\delta & \sin\delta \\ -\sin\delta & \cos\delta\end{array} \right),
\end{equation}
where $\cos\delta\propto 1/\sqrt{\log N}$.

The new evolution operator is
\begin{equation}\label{eq:new_U}
    U''=(-Z\otimes I)\cdot C(U)\cdot (X_{\delta}^{\dagger}\otimes I) \cdot C(R_{\hat t}) \cdot (X_{\delta}\otimes I),
\end{equation}
where $C(U)$ and $C(R_{\hat t})$ are the controlled operations shown in Figure~\ref{fig:circ} and $I$ is the identity operator in ${\cal H}$. We will show that $U''$ must be iterated $O(\sqrt{N \log N})$ times, taking $\ket{1}\ket{u_C}\ket{u_P}$ as the initial condition, in order to maximize the overlap with the search element.

The expression for the controlled $R_{\hat t}$ is $C(R_{\hat t})=I_{12N}-2\ket{1,u_C,\hat t}\bra{1,u_C,\hat t}$. Let us define
\begin{equation}
    \ket{\delta_1} \equiv X_\delta^\dagger\ket{1}=-\sin \delta \, \ket{0} + \cos \delta \, \ket{1},
\end{equation}
then we define a new reflection operator
\begin{equation}
    {\bar R}_{\hat t}\equiv(X_{\delta}^{\dagger}\otimes I) \cdot C(R_{\hat t}) \cdot (X_{\delta}\otimes I) = I_{12N} - 2\ket{\delta_1,u_C,\hat t}\bra{\delta_1,u_C,\hat t}.
\end{equation}
The effective target state is
\begin{equation}
    \ket{{\bar t}}=\ket{\delta_1,u_C,\hat t}.
\end{equation}

Let us define
\begin{equation}
    \bar U=(-Z\otimes I)\cdot C(U).
\end{equation}
Note that the eigenspectrum of $\bar U$ is determined from the eigenspectrum of $U$. In fact, for $\hat k\neq 0$
\begin{eqnarray}
    {\bar U}\ket{0}\ket{\nu_{\pm k},\hat k} &=& -\ket{0}\ket{\nu_{\pm k},\hat k},\\
    {\bar U}\ket{1}\ket{\nu_{\pm k},\hat k} &=& \textrm{e}^{\pm i\theta_k} \ket{1}\ket{\nu_{\pm k},\hat k}.
\end{eqnarray}
Eigenvectors $\ket{l}\ket{\nu_{\pm 1},\hat k}$, $l=0,1$  will not be used, because $\scalar{u_C}{\nu_{\pm 1}}=0$. They are orthogonal both to the initial condition and to the target. For $\hat k=0$, the initial condition is the only eigenvector with eigenvalue 1 that will be used.

The search algorithm consists in applying $U''$ repeatedly taking the initial condition as
\begin{equation}\label{initial_condition}
    \ket{\Psi_0}\equiv\ket{1,u_C,u_P}=\frac{1}{\sqrt{6\,N}}\sum_{j,\hat n} \ket{1,j,\hat n},
\end{equation}
where $\ket{u_C}$ and $\ket{u_P}$ are the uniform superposition in the coin and position spaces respectively. The initial condition can be prepared in $O(\sqrt N)$ time steps. The number of iterations of $U''$ is given by $\pi/4\alpha$, where $\alpha$ is defined in the following way. The eigenvalues of $U''$ that are different from 1 have the form $\textrm{e}^{\pm i \alpha_j}$, where $0< \alpha_j\le \pi/2$. Then $\alpha=\min\{\alpha_1,\alpha_2,\cdots\}$. AKR have shown that it is possible to estimate $\alpha$ knowing the eigenspectrum of the evolution operator $\bar U$.

The first step in the analysis of the algorithm is to decompose the target state in the eigenbasis of $\bar U$, that is, we have to calculate coefficients $a_0$, $a_{k}$ and $b_k$ such that
\begin{eqnarray}\label{target_decomposition}
    \ket{\delta_1,u_C,\hat t} &=& \frac{a_0}{\sqrt N}\ket{\Psi_0} + \frac{1}{\sqrt N}\sum_{\hat k\neq 0} a_{k}\left(\ket{1,\nu_{+k},\hat k}+\ket{1,\nu_{-k},\hat k}\right) + \nonumber\\
    &&\frac{1}{\sqrt N}\sum_{\hat k\neq 0} b_{k}\left(\ket{0,\nu_{+k},\hat k}+\ket{0,\nu_{-k},\hat k}\right).
\end{eqnarray}
Coefficients $b_k$ are related to the $(-1)$-eigenspace. Coefficients $a_{k}$ and $b_k$ are real. In order to satisfy these conditions, eigenvectors $\ket{l,\nu_{+k},\hat k}$ for $l=0,1$ must be chosen appropriately. The procedure is the following. If $\scalar{u_C}{{\nu_{+k}}}$ has a non-zero phase $\textrm{e}^{i\lambda}$, then $\ket{l,\nu_{+k},\hat k}$ must be redefined to $\textrm{e}^{-i\lambda}\ket{l,\nu_{+k},\hat k}$. The same procedure must be performed to $\ket{l,\nu_{-k},\hat k}$. After those redefinitions, Eq.~(\ref{target_decomposition}) is valid. For the triangular network, a straightforward calculation yields
\begin{eqnarray}
    a_{k}&=&\scalar{1}{\delta_1}\scalar{\nu_{{\pm}k}}{u_C}= \frac{\cos \delta}{\sqrt 2}\label{a_k_triangle}\\
    b_{k}&=&\scalar{0}{\delta_1}\scalar{\nu_{{\pm}k}}{u_C}= -\frac{\sin \delta}{\sqrt 2}\label{b_k_triangle}
\end{eqnarray}
This result is remarkably similar to the result obtained for the 2D grid. Note that the corresponding expressions for the honeycomb lattice~\cite{Hexagons} are more complex, because they depend on $k$.

Tulsi has shown that $\alpha$ can be determined using the expression
\begin{equation}\label{eq:AN}
  \frac{1}{\alpha}=O\left(\frac{1}{a_0}\sqrt{\sum_{\hat k\neq {0}}\frac{a_{k}^2}{1-\cos\theta_k}+\sum_{\hat k\neq 0} \frac{b_k^2}{4}}\right),
\end{equation}
when $N\gg 1$. Eqs.~(\ref{a_k_triangle}) and (\ref{b_k_triangle}) show that $a_k$ and $b_k$ do not depend on $\hat k$. The non-trivial sum inside the above square root may be calculated using
\begin{equation}\label{eq:integra1}
  \frac{1}{2}\sum_{{k}\neq 0}\frac{1}{1-\cos\theta_k}\approx\frac{N}{16}
\frac{1}{(\pi-\varepsilon)^2} \iint_\varepsilon^{2\pi-\varepsilon}d\tilde k_2 d\tilde k_1 \frac{1}{1-\cos\theta_k},
\end{equation}
where we have used $\sum_{{k}\neq 0}\rightarrow\frac{N}{8}\frac{1}{(\pi-\varepsilon)^2}\iint_\varepsilon^{2\pi-\varepsilon}
d\tilde k_1 d\tilde k_2$ with $\varepsilon=2\pi\sqrt{2/N}\ll 1$. Using Eq.~(\ref{cos_theta}) and $\varepsilon\ll 1$, Eq.~(\ref{eq:integra1}) can be approximated by
\begin{equation}\label{eq:timecompl}
\frac{3N}{16}\frac{1}{\pi^2}\int_\varepsilon^{2\pi-\varepsilon}
d\tilde k_2 \int_\varepsilon^{2\pi-\varepsilon} \frac{d\tilde k_1}{\tilde k_1^2 + \tilde k_2^2 - \tilde k_1\tilde k_2}
\sim N\log\left(\frac{2\pi}{\varepsilon}\right)\sim N\log N.
\end{equation}
Replacing this result in Eq.~(\ref{eq:AN}), using that $a_0=\cos\delta$, $\cos\delta=\Theta\left(1/\sqrt{\log N}\right)$ and Eqs.~(\ref{a_k_triangle}) and (\ref{b_k_triangle}), we obtain
\begin{equation}
    \frac{1}{\alpha}=O\left(\sqrt{N\log N}\right).
\end{equation}
Since the number of iterations of $U''$ is $\pi/4\alpha$, we conclude that the time complexity of the search algorithm is $O(\sqrt{N\log N})$. It remains to show that the probability to find the marked vertex is constant. This is the moment where Tulsi's method shines.

The overlap of the final state after $t=\pi/4\alpha$ iterations of $U''$ with the target state is
\begin{equation}
    |\bra{\delta_1,u_C,\hat t}(U'')^t\ket{\Psi_0}|=\Theta\left(\min\left\{\frac{1}{\sqrt{\sum_{\hat k\neq 0}a_k^2\cot^2\frac{\theta_k}{4}}},1\right\}\right).
\end{equation}
The coefficients $b_k$ play no role in this overlap as shown by Tulsi. The calculation at this point for the triangular network is similar to one performed by AKR for the 2D grid, except that $a_k$ has the constant factor of $\cos\delta$. Using that: (1) $\cot \theta_k\leq \frac{1}{\theta_k^2}$, (2) $\frac{1}{\theta_k^2}$ is bounded from below and above by \textit{const}$/(1-\cos \theta_k)$ and (3) the result of Eq.~(\ref{eq:timecompl}), we obtain
\begin{equation}
    |\bra{\delta_1,u_C,\hat t}(U'')^t\ket{\Psi_0}|=\Theta\left(\min\left\{\sqrt{\cos^2(\delta) \log N},1\right\}\right).
\end{equation}
Using that  $\cos\delta=\Theta\left(1/\sqrt{\log N}\right)$, we obtain
\begin{equation}
    |\bra{\delta_1,u_C,\hat t}(U'')^t\ket{\Psi_0}|=\Theta\left(1\right).
\end{equation}
The probability of finding the marked vertex is constant. We have confirmed this result with a numerical simulation in Fig.~\ref{fig:overlap}.

\section{Simulation of the search algorithm}
\label{sec:simulation}

A generic state in the extended space is
\begin{equation}
 \ket{\Psi}=\sum_j \sum_{\hat n} a_{j,\hat{n}}\, \ket{0, j,\hat{n}} + b_{j,\hat{n}}\, \ket{1, j,\hat{n}}, \label{eq:ext_state_vector}
\end{equation}
where $a,b$ are the amplitudes of the state vector extended to include the 2-dimensional Hilbert space associated with the ancilla qubit.

It is straightforward to show that the operator $U''$ in Eq. (\ref{eq:new_U}) defines the following map:
\begin{eqnarray}
 \tilde a_{j,\hat{n}} &=& -a_{j,\hat{n}} + \frac13\delta_{\hat{n},\hat{t}}  \sum_{\ell}\left( a_{\ell,\hat{t}}\sin^2 \delta - b_{\ell,\hat{t}}\cos \delta \sin \delta \right), \\ \nonumber
 \tilde b_{j,\hat{n}} &=& \sum_{k,l} G_{k,l}^{(6)} \sum_{\hat{n}^\prime} S_{j,k,\hat{n},\hat{n}^\prime} \left[ b_{l,\hat{n}^\prime} + \frac13\delta_{\hat{n}^\prime,\hat{t}} \sum_\ell \left( a_{\ell,\hat{t}}\sin \delta \cos \delta
- b_{\ell,\hat{t}}\cos^2 \delta \right)
\right]\nonumber
\end{eqnarray}
where $\delta_{\hat n,\hat t}$ is a Kronecker-delta which selects the searched state. This map allows the simulation of the abstract search algorithm in a digital computer. The initial condition is taken to be $ \ket{1}\ket{u_C}\ket{u_P} $, and the effective target state is
$\ket{\bar{t}}=\ket{\delta_1}\ket{u_C}\ket{\hat{t}}$, as explained in the last section.

Fortran 90 was chosen as programming language because it provides useful tools for large matrix computations and intrinsic functions for complex vector algebra. A parallel programming approach using OpenMP was implemented.

\begin{figure}[h]
  \centering
  \includegraphics[scale=.24, angle=-90]{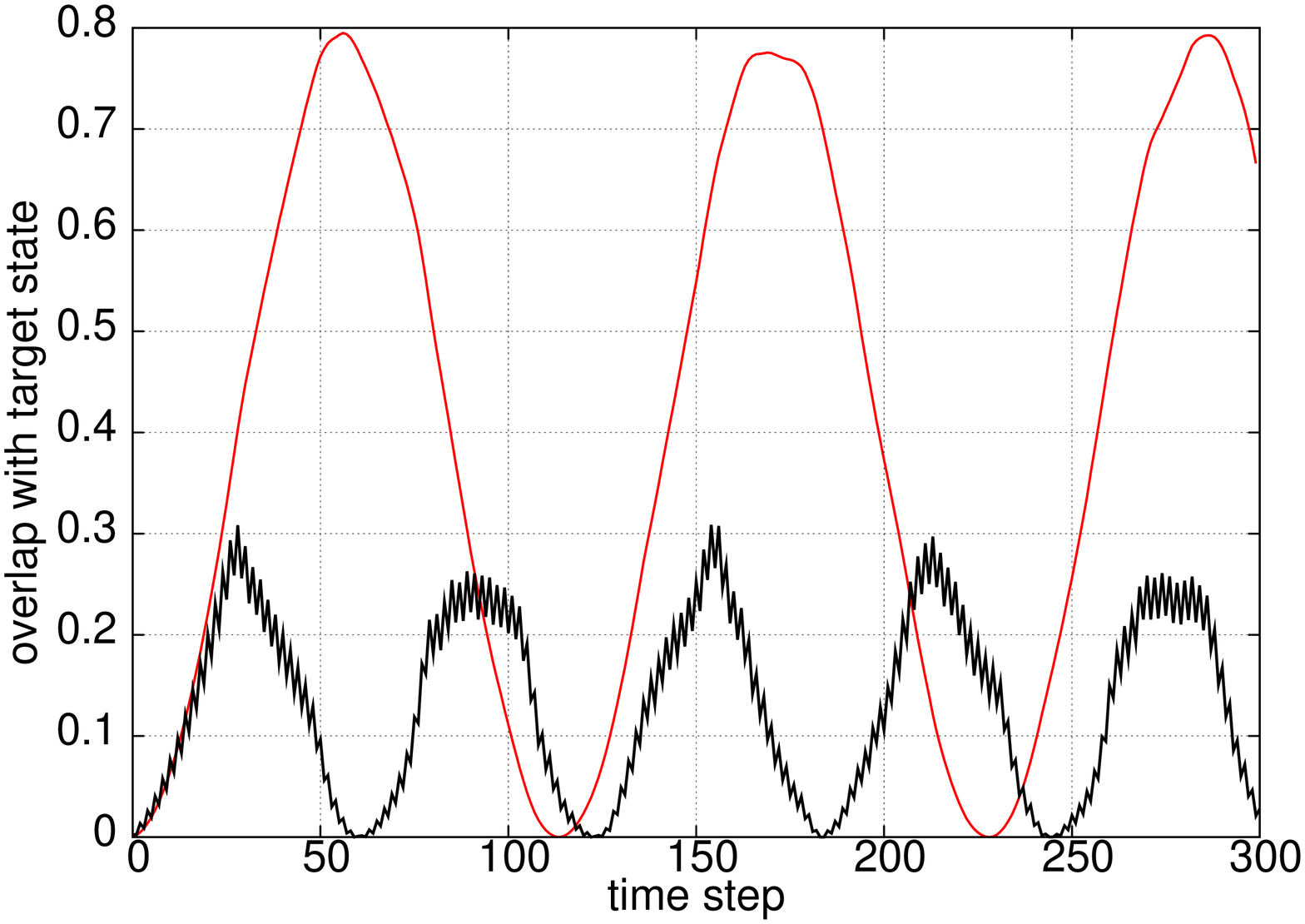}~
  \includegraphics[scale=.24, angle=-90]{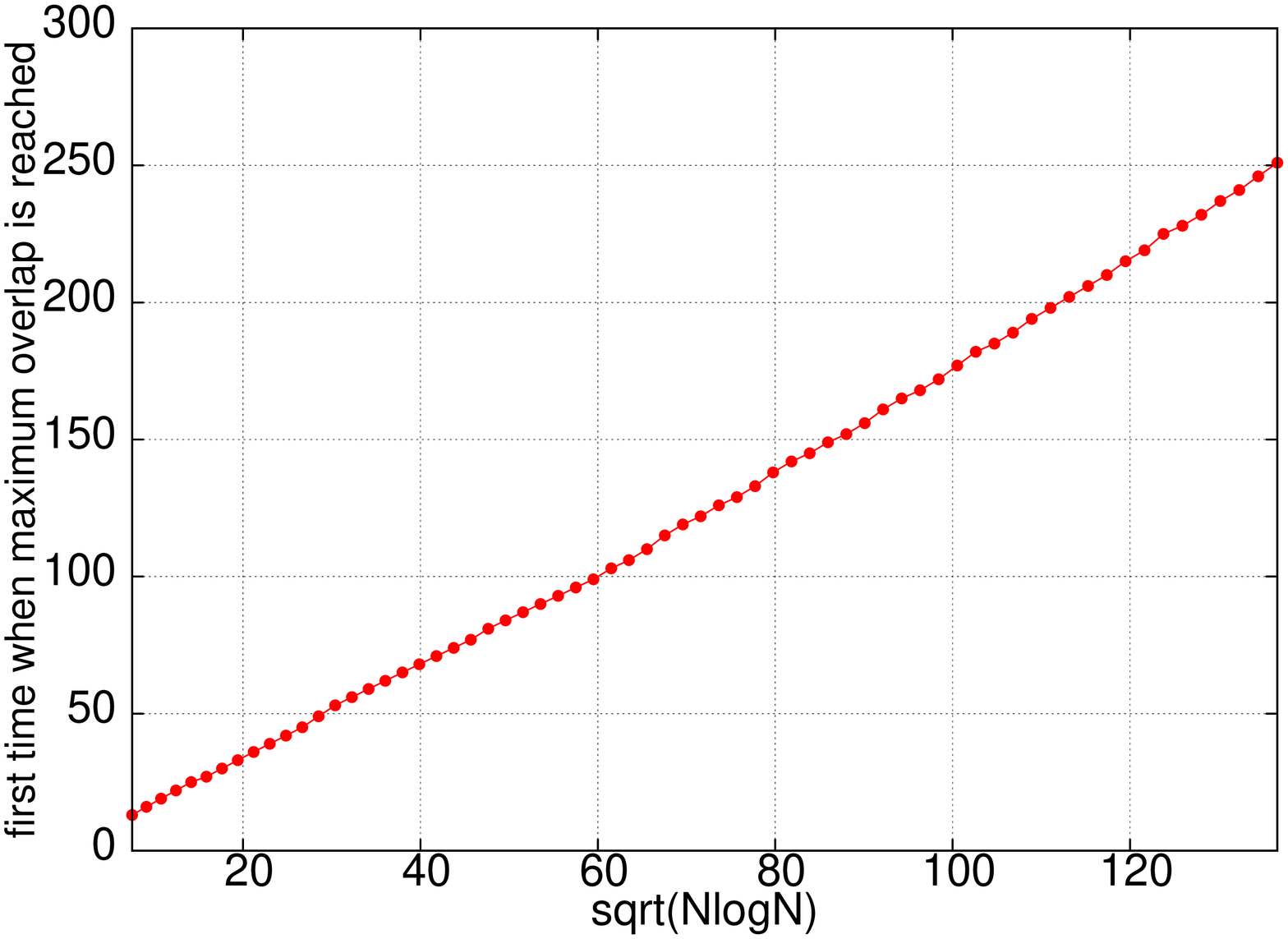}
  \caption{Left panel: probability of finding the searched state against the number of time steps for spatial search with (thin line) and without (thick line) Tulsi's modification for $N=400$. Right panel: $t_{\rm{MAX}}$ against $\sqrt{N\log N}$ using Tulsi's search. Note the approximately linear dependence. }
\label{fig:Tpm}
\end{figure}

\begin{figure}[h]
  \centering
  \includegraphics[scale=.3, angle=-90]{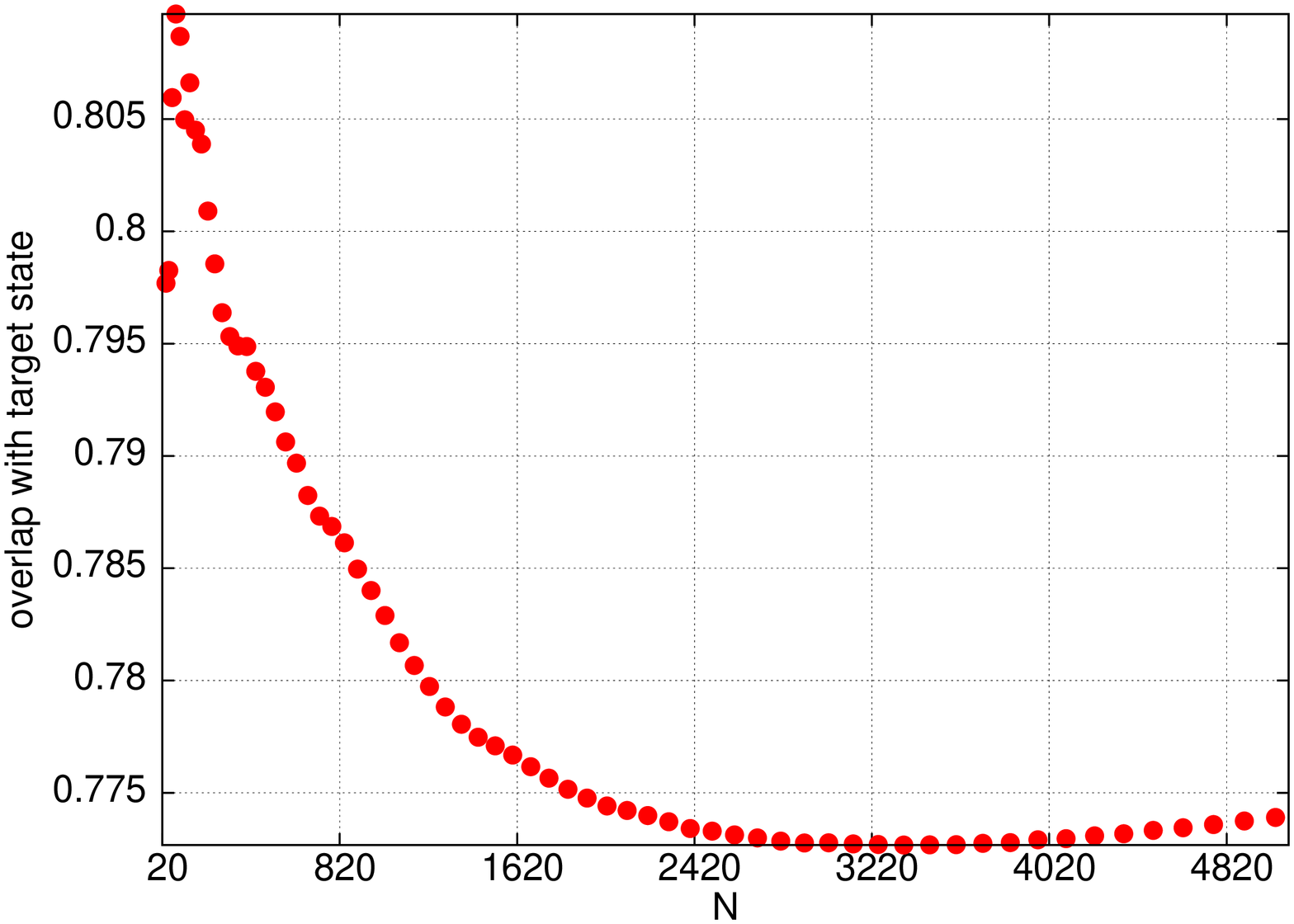}
  \caption{Probability of finding the marked vertex $\big|\bra{{\bar t}}{(U'')^{t_{max}}}\ket{1,u_C,u_P}\big|^2$ against $N$.}
\label{fig:overlap}
\end{figure}

We present the results of the simulations in Figures~\ref{fig:Tpm} and \ref{fig:overlap}.  The left panel in Figure~\ref{fig:Tpm} shows the time evolution of the probability of finding the searched state, $\big|\scalar{\bar t}{\Psi}\big|^2$, both for an algorithm with Tulsi's modification (thin curve), based on $U''$ defined in Eq.~(\ref{eq:new_U}), and without it (thick curve), based on $U'=S\cdot C'$ with the modified coin operator defined in Eq.~(\ref{coin_operator}). Note that Tulsi's search yields a smoother curve, the first maximum $(t_{max})$ of which is reached later but the maximum probability of finding the searched element is higher than with the usual spatial search. A change in the sign of the local derivative was used to find the maximum point. In the right panel of Figure~\ref{fig:Tpm}, the time at which the maximum probability is reached for Tulsi's search is plotted against $\sqrt{N\log N}$. A straight-line fit to this data has a correlation coefficient $R^2=0.9988$.

In order to test whether the overlap with the searched site in Tulsi's search is constant for large $N$, in Figure~\ref{fig:overlap} we plot this overlap, $\big|\bra{{\bar t}}{(U'')^{t_{max}}}\ket{1,u_C,u_P}\big|^2$, against $N$.
After an initial decay, for $N>2000$ the overlap is found to stabilize at approximately $0.773$. These simulations show that Tulsi's search in a triangular network works as expected.

\section{Summary and conclusions}
\label{sec:final}

In this work the problem of quantum spatial search on a periodic two-dimensional triangular network has been considered for the first time. In this problem, a searched item is to be located in a regular triangular network with $N$ elements. A quantum walk operator for this network has been defined and its explicit Fourier representation has been found. Its eigenproblem was solved exactly and these results have been used to estimate the overlap and running times of the algorithm, according to the generalized search formalism  \cite{AKR05}. This formalism gives a powerful insight on the runtime of the algorithm for large values of $N$. In order to check our analytical results and to gain further knowledge on the detailed performance of the spatial search, we have implemented its simulation on a classical computer, using standard parallel techniques. This allowed us to obtain results for high values of $N$ and see how fast the convergence to the theoretical expectations actually is. The simulations were implemented both for a modified quantum walk search and for Tulsi's search, which uses an extra qubit as a control register.

Both spatial search algorithms are found to require $O(\sqrt{N\log N})$ steps to reach the point at which a measurement yields the searched state with constant probability.
However, this result is obtained with a constant overlap with the searched state, in the case of Tulsi's search.

Previous work for the spatial search problem in a plane has considered the case of a square grid \cite{AKR05} and an hexagonal grid \cite{Hexagons}. For both quantum algorithms the time complexity is  $O(\sqrt{N\log N})$. In a sense, this work completes the program for the spatial search problem, by providing the details of a search in a triangular grid. Since these regular graphs have different degree ($d=4$ for rectangles, $d=3$ for hexagons and $d=6$ for triangles), this completes the proof that the degree of a regular graph does not affect the performance of a spatial search algorithm.

A search algorithm implemented on a real network will have to cope with loss of symmetry due to imperfections. The issue of how robust these different spatial search protocols are when it comes to searching when a fraction of the links in the network is missing is still an important open question which remains for future work.

\section*{Aknowledgements}  This work was done with financial support from PEDECIBA (Uruguay) and CNPq (Brazil).


\end{document}